\documentstyle[sprocl]{article}

\def\eq{\begin{eqnarray}}
\def\en{\end{eqnarray}}

\begin{document}

\title{THEORY OF HADRONIC ATOMS\footnote{
Present work is based on the results obtained in collaboration with
A.~Gall, J.~Gasser, V.E.~Lyubovitskij and E.~Lipartia}}

\author{A. RUSETSKY}

\address{Institute for Theoretical Physics, University of Bern,\\
Sidlerstrasse 5, 3012 Bern, Switzerland\\E-mail: rusetsky@itp.unibe.ch}

\maketitle\abstracts{
A review of the modern theory of hadronic atoms based on the non-relativistic
effective Lagrangian approach and ChPT, is given. 
As an application of the general framework, we consider the decay of
the $\pi^+\pi^-$ atom into two neutral pions, and the energy-level
shift of the $\pi^-p$ atom. The analysis of the decay width of the
$\pi^+\pi^-$ atom has been carried out at the first non-leading 
order in isospin breaking, and at $O(e^2p^2)$ in ChPT, that results in
an accurate theoretical prediction for this quantity. 
The study of the $\pi^-p$ atom energy-level shift within the same approach
clearly demonstrates the necessity to critically reaccess
the values of the $\pi N$ scattering lengths, extracted from the energy-level
shift measurement by means of the potential model-based theoretical
analysis. The construction of short-range hadronic potentials from
field theory is also discussed.    
}

\section{Introduction}

Recent years have seen a  growing interest in the study of hadronic bound
states - so called hadronic atoms -
that provides an extremely valuable piece of information on the behavior
of QCD at a very low energy.  
At CERN, the DIRAC collaboration~\cite{DIRAC} aims to measure the 
$\pi^+\pi^-$ atom lifetime to $10\%$ accuracy. This would allow one to 
determine the difference $a_0-a_2$ of $\pi\pi$ scattering lengths with $5\%$
precision. This measurement provides a crucial test for the large/small
condensate scenario in QCD: should it turn out that the quantity $a_0-a_2$ 
is different from the value predicted in standard ChPT~\cite{ChPT}, one has 
to conclude~\cite{Stern} that spontaneous  chiral symmetry breaking in QCD 
proceeds differently from the widely accepted picture. In the experiment 
performed at PSI~\cite{PSI1,PSI2}, one has measured the strong energy-level 
shift and the total decay width of the $1s$ state of pionic hydrogen, as 
well as the $1s$ shift of pionic deuterium. These measurements  yield 
isospin symmetric $\pi N$ scattering lengths to an accuracy which is unique 
for hadron physics. A new experiment on  pionic hydrogen at PSI has recently
been  approved. It will allow one to measure the decay 
$A_{\pi^-p}\rightarrow \pi^0n$  to much higher accuracy and thus enable one, 
in principle, to determine the $\pi N$ scattering lengths from  data on 
pionic hydrogen alone. This might vastly reduce the model-dependent 
uncertainties that come from the analysis of the three-body problem in
$A_{\pi^-d}$. Finally, the DEAR collaboration~\cite{DEAR} at the DA$\Phi$NE 
facility plans to measure the energy level shift and lifetime of the $1s$ 
state in $K^{-}p$ and $K^-d$ atoms - with considerably higher precision 
than in the previous experiment carried out at KEK~\cite{KEK} for
$K^-p$ atoms. It is expected~\cite{DEAR} that this will result in a 
precise determination of the $I=0,1$ $S$-wave  scattering lengths.
It will be a challenge for theorists to extract from this new information 
on the $\bar{K}N$ amplitude at threshold a more precise value of e.g. the 
isoscalar kaon-sigma term and of the strangeness content of the nucleon.

In order to carry out the precision experimental tests of QCD mentioned 
above, on the theoretical side one faces the problem of finding the
suitable field-theoretical framework for the description of the
measured characteristics of hadronic atoms - energy levels and decay 
probabilities. In this work we shall report on the recent progress 
in this direction. The main message which will be delivered, is that a 
rigorous theory of this sort of bound states based on the merger of ChPT
and the non-relativistic effective Lagrangian technique that was proposed
originally by Caswell and Lepage~\cite{Lepage} to study QED bound states in
general, can be indeed constructed. This theory allows for a systematic
expansion of the bound-state observables both in quark mass 
$\hat m=\frac{1}{2}\,(m_u+m_d)$, and the
isospin-breaking parameters: fine structure constant $\alpha$ and the quark
mass difference $m_d-m_u$ (for simplicity, we consider an $SU(2)$ case here).

Hadronic atoms appear to be loosely bound systems of hadrons that are formed
mainly by the static Coulomb force. The Bohr radius of this sort of bound
states is of order of a few hundreds of Fm, and the average 3-momenta of
constituents lie in MeV range. For the above reason, it is evident that
the non-relativistic framework provides the most natural and economical tool
for handling such bound states. Relativistic corrections are taken into
account perturbatively, up to any given order in the expansion in inverse
powers of masses. Below, we briefly outline the main ingredients of the
non-relativistic effective Lagrangian approach to bound states, without going
into details.  

$\bullet$ The non-relativistic Lagrangian describing interactions between
hadrons and photons, is built from the non-relativistic hadron fields and the
photon field. This Lagrangian consists of an infinite tower of all possible
operators, with an increasing mass dimension - all operators allowed by
discrete symmetries and the gauge invariance should be included, with an 
{\it a priori} unknown couplings. In actual calculations, only a
few low-dimensional ones matter - higher-dimensional operators contribute to
higher powers in $\alpha$ in the bound-state observables. This feature of the
effective theory goes under the name of ``power counting''. Further, the
Lagrangian does not include, by definition, the operators that change the
number of heavy particles (hadrons).

$\bullet$ Loop corrections to the scattering amplitudes in the 
non-relativistic theory are calculated in a standard manner, by using the 
Feynman diagrammatic technique. There is, however, one important modification.
It is well known that in the non-relativistic theory in the 
presence of light particles (photons) the Feynman
integrals should be properly regularized in order to avoid the contribution
from the loop momenta at the hard scale - otherwise, loop corrections to
the Green functions would lead the the breakdown of counting rules in the
non-relativistic theory. A suitable regularization procedure built on the top
of the Feynman rules in the non-relativistic theory is provided
by so-called ``threshold expansion''~\cite{Beneke}, that enables one to
disentangle the contributions coming from different regions of loop momenta,
by expanding the integrands - in the dimensional regularization - in all
possible small kinematical variables. Next, one has to systematically remove
hard-momentum contribution from the integrals, which at low energies 
is given by a polynomial in external momenta, and can be absorbed into the
renormalization of the couplings in the non-relativistic Lagrangian.
For a more detailed discussion of the problem, see~\cite{Lamb}. 

$\bullet$ Couplings in the non-relativistic effective Lagrangian are
determined from the matching to the relativistic theory. These couplings are
not necessarily real, since the physical decay processes where the number of
the hadrons is not conserved, contribute to the imaginary parts of these
couplings. The matching condition determines the couplings in terms of
threshold parameters of the physical hadronic scattering amplitudes. It is
crucial to stress that the matching condition does not imply the chiral
expansion of the amplitudes - formally, it is valid in all orders in the
chiral expansion.

$\bullet$ After setting the parameters in the non-relativistic Lagrangian, we
turn to the bound states in the theory. The Feshbach
formalism~\cite{Feshbach} that allows one to separate the bound-state pole in
the scattering amplitudes, turns out to be very convenient for this purpose.
The real and imaginary parts of the pole position on the second Riemann sheet
of the complex energy plane coincide, by definition, with the energy and the
decay width of the (metastable) bound state. The perturbative framework for
determining the pole position coincides with the conventional
Rayleigh-Schr\"{o}dinger perturbation theory. At the end, the characteristics
of the bound state are determined through the couplings in the
non-relativistic amplitudes. With the use of the matching condition, these
characteristics can be further expressed via the scattering amplitudes in the
relativistic theory.

$\bullet$ Given the chiral expansion for the scattering amplitude, it is
possible to obtain the chiral expansion for the bound-state observables which
can be reorganized in the expansion in $\hat m$ and isospin-breaking
parameters. The isopin symmetry world is, by convention, defined as the one
where the masses of pions an nucleons coincide with the charged ones, and the
values of all low-energy constants remain the same. In this vein, one may
achieve an unambiguous separation of the isospin-breaking effects and extract
the isospin-symmetric hadronic scattering lengths directly from the hadronic
atom measurements.

After having described the general framework, we consider to the particular
systems.  

\section{Decay of the $\pi^+\pi^-$ atom into $\pi^0\pi^0$}

Recently, using the non-relativistic effective Lagrangian framework,
a general expression for the decay width $\Gamma_{A_{2\pi}\to\pi^0\pi^0}$
of the $1s$ state of the $\pi^+\pi^-$ atom  was obtained at 
next-to-leading order in isospin-breaking~\cite{Bern1}. Numerical analysis
of this quantity was carried out at order $O(e^2p^2)$ in ChPT~\cite{Bern2}.
These investigations have confirmed and generalized  the results of earlier 
studies~\cite{Sazdjian,Dubna}. The expression for the decay width at the first
non-leading order in isospin breaking has the form
\eq\label{general-pipi}
&&\Gamma_{A_{2\pi}\to\pi^0\pi^0}=\frac{2}{9}\,\alpha^3 p^\star 
{\cal A}_{\pi\pi}^{~2} (1+K_{\pi\pi})\, ,
\\[2mm]
&&K_{\pi\pi}=
\frac{\Delta M_\pi^2}{9M^2_{\pi^+}}\,(a_0+2a_2)^2
-\frac{2\alpha}{3}\,(\ln\alpha-1)(2a_0+a_2)+o(\alpha,(m_d-m_u)^2)\, . 
\nonumber
\en
Here $p^\star=(M_{\pi^+}^2-M_{\pi^0}^2-\frac{1}{4}M_{\pi^+}^2\alpha^2)^{1/2}$,
and $a_I$, $(I=0,2)$ denote the (dimensionless) strong $S$-wave
$\pi\pi$ scattering lengths
in the channel with total isospin $I$. The quantity ${\cal A}_{\pi\pi}$ is
obtained as follows~\cite{Bern1}. One calculates the relativistic amplitude
for the process $\pi^+\pi^-\rightarrow\pi^0\pi^0$ at $O(\alpha,(m_d-m_u)^2)$
in the normalization chosen so that at $O(1)$ the amplitude at threshold 
coincides with the difference $a_0-a_2$.
Due to the presence of virtual photons,
the amplitude is multiplied by an overall Coulomb phase $\theta_c$ 
that is removed. The real part of the remainder contains terms
that diverge like $|{\bf p}|^{-1}$ and $\ln 2|{\bf p}|/M_{\pi^+}$ at
$|{\bf p}|\rightarrow 0$ (${\bf p}$ denotes the relative 3-momentum
of charged pion pairs). The quantity ${\cal A}_{\pi\pi}$
 is obtained by subtracting these divergent pieces, and by then 
evaluating the remainder at ${\bf p}=0$.
\eq\label{threshold-pipi}
{\rm Re}\,({\rm e}^{-i\theta_c}\, t_{\pi\pi})&\rightarrow&
\frac{b_1}{|{\bf p}|}+b_2\ln\frac{2|{\bf p}|}{M_{\pi^+}}+
\frac{8\pi}{3M_{\pi^+}^2}\,{\cal A}_{\pi\pi}+\cdots
\en

As it is seen explicitly from Eq.~(\ref{general-pipi}), one can directly 
extract the value of ${\cal A}_{\pi\pi}$ from the measurement of the  
decay width, because the correction  $K_{\pi\pi}$ is very small and the 
error introduced by it is negligible. In order to extract strong scattering
lengths from  data, one may invoke ChPT
and to relate the quantities ${\cal A}_{\pi\pi}$ and
$a_0-a_2$ order by order in the chiral expansion~\cite{Bern2}. This requires
the evaluation of isospin-breaking corrections to the scattering amplitude.
At order $O(e^2p^2)$ in chiral expansion we obtain
\eq\label{numerics}
A_{\pi\pi}=a_0-a_2+\epsilon_{\pi\pi}\, ,\quad
\epsilon_{\pi\pi}=(0.58\pm 0.16)\cdot 10^{-2}\, ,\quad
K_{\pi\pi}=1.15\cdot 10^{-2}\, .
\en
The lifetime of the $\pi^+\pi^-$ atom is predicted to be
\eq
\tau_{2\pi^0}=(2.91\pm 0.09)\cdot 10^{-15}~{\rm s}\, ,
\en
and the correction to the leading-order in isospin-breaking
$\Gamma_{2\pi^0}=\Gamma^{LO}_{2\pi^0}(1+\delta_\Gamma)$ where
$\Gamma^{LO}_{2\pi^0}=\frac{2}{9}\,\alpha^3 p^\star(a_0-a_2)^2$, equals to
$\delta_\Gamma=0.056\pm 0.012$. 
Note that above we have used the values of $\pi\pi$ scattering lengths
$a_0=0.220,~a_2=-0.0444,~\Delta(a_0-a_2)=0.004$, obtained on the basis of
two loop calculations in ChPT and dispersion relations
analysis~\cite{Colangelo}. For the various low-energy constants
entering the expression for $\epsilon_{\pi\pi}$, the same values as in
Ref.~\cite{Bern2} were used. These values of the low-energy constants 
will not necessarily coincide with the ones that will emerge from the analysis
of scattering lengths.
In addition, the errors in the scattering lengths and the
correction term at this stage are treated as uncorrelated. 
Only once the new values for 
the low-energy constants $\bar l_i$ from the dispersion analysis are available
including error bars, one may refine the above preliminary analysis for the 
decay width. We also emphasize,
that the corrections to the decay width at $O(e^2p^2)$ are already very
small, justifying the negligence of the higher-order terms~\cite{Bern2}.

\section{Energy level shift of the $\pi^- p$ atom}

The treatment of the $\pi^- p$ atom problem~\cite{Bern3} proceeds along the 
lines very similar to those for $\pi^+\pi^-$ case. Our investigations
are aimed at the derivation of the general expression for the $\pi^- p$
atom energy-level shift in the $1s$ state. The total shift is given by a sum
of the electromagnetic and strong pieces. Our calculations for
the electromagnetic shift~\cite{Bern3} within a high accuracy yield the same 
result as given
in Ref.~\cite{PSI1}. The final result for the strong shift $\epsilon_{1s}$ 
in the first non-leading order in isospin breaking is given in a form similar 
to Eq.~(\ref{general-pipi})
\eq\label{general-piN}
&&\epsilon_{1s}=-2\alpha^3\mu_c^2\, {\cal A}_{\pi N}\,(1+K_{\pi N})
\\[2mm]
&&K_{\pi N}=-2\alpha\mu_c(\ln\alpha-1){\cal A}_{\pi N}+o(\alpha,m_d-m_u)\, ,
\nonumber
\en
where $\mu_c$ denotes the reduced mass of the $\pi^- p$ pair, and the quantity
${\cal A}_{\pi N}$ is defined analogously to ${\cal A}_{\pi\pi}$. To calculate
this quantity, one has to evaluate the $\pi^- p\rightarrow\pi^- p$ 
relativistic scattering amplitude at $O(\alpha,m_d-m_u)$, drop all 
diagrams that are made disconnected by cutting one photon line, and
discard the spin-flip piece. The remainder
is denoted by $t_{\pi N}$. The regular part of $t_{\pi N}$ at
threshold defines the quantity ${\cal A}_{\pi N}$ in analogy to
Eq.~(\ref{threshold-pipi})
\eq\label{threshold-piN}
{\rm Re}\,({\rm e}^{-2i\theta_c}\, t_{\pi N})\rightarrow
\frac{B_1}{|{\bf p}|}+B_2\ln\frac{|{\bf p}|}{\mu_c}-
\frac{2\pi}{\mu_c}\,{\cal A}_{\pi N}+\cdots\, .
\en
  
In order to extract the value of the $S$-wave $\pi N$ scattering lengths
$a^+_{0+},~a^-_{0+}$ from the $\pi^- p$ energy shift measurement, one
may again resort to ChPT, to calculate the isospin-breaking corrections
to the $\pi N$ scattering amplitude at threshold. 
The normalization of the relativistic amplitude 
is chosen so that at threshold 
${\cal A}_{\pi N}=a^+_{0+}+a^-_{0+}+O(\alpha,m_d-m_u)$.
We have carried out these calculations at chiral order $O(p^2)$,
where only the tree diagrams contribute, the result looks as follows 
\eq\label{amplitude}
{\cal A}_{\pi N}=a^+_{0+}+a^-_{0+}+\epsilon_{\pi N}\, ,\quad\quad
\epsilon_{\pi N}=\frac{m_p(8c_1\Delta M_\pi^2-4e^2f_1-e^2f_2)}
{8\pi(m_p+M_{\pi^+})F^2}\, , 
\en
where $c_i$ ($f_i$) are the strong (electromagnetic) low-energy constants 
from the $O(p^2)$ Lagrangian of ChPT~\cite{Lagrangian}.
In order to perform the numerical analysis,
one has to specify the values of these low-energy constants. The "strong" 
constant
$c_1$ can be determined from the fit of the elastic $\pi N$ scattering
amplitude at threshold to KA86 data~\cite{private}: 
$c_1=-0.925~{\rm GeV}^{-1}$. The value of the constant $f_2$
can be extracted from the proton-neutron electromagnetic mass
difference~\cite{Reports}: $e^2f_2=(-0.76\pm 0.3)~{\rm MeV}$. The 
determination of the constant $f_1$ from data is however, problematic. For
this reason, in our analysis we have used order-of-magnitude estimate for 
this constant: $-|f_2|\leq f_1\leq |f_2|$. With these values of the low-energy
constants, we obtain the isospin-breaking correction to the leading-order
energy-level shift defined as
$\epsilon_{1s}=\epsilon_{1s}^{LO}(1+\delta\epsilon)$, where
$\epsilon^{LO}_{1s}=-2\alpha^3\mu_c^2(a^+_{0+}+a^-_{0+})$, to be
$\delta_\epsilon=(-4.8\pm 2.0)\cdot 10^{-2}$.
The large uncertainty is caused mainly by the poor knowledge of the parameter
$f_1$. In demonstration of the above discussion, in Fig.~1 we confront the
results of the above analysis with those of the potential model~\cite{PSI2}:
$\delta_\epsilon=(-2.1\pm 0.5)\cdot 10^{-2}$,
using the same experimental input. As it is readily seen from Fig.~1, the
systematic error in the potential approach is grossly underestimated,
that already indicates at the necessity to critically reaccess the values of
the $S$-wave $\pi N$ scattering lengths quoted in Ref.~\cite{PSI2}.
In addition, it remains to be seen, how our results will be altered by the
loop corrections in ChPT. A precise determination of the low-energy constant
$f_1$, using either sum rules or invoking various models, is also desirable.

\begin{figure}[t]
 \vspace{7.0cm}
\includegraphics{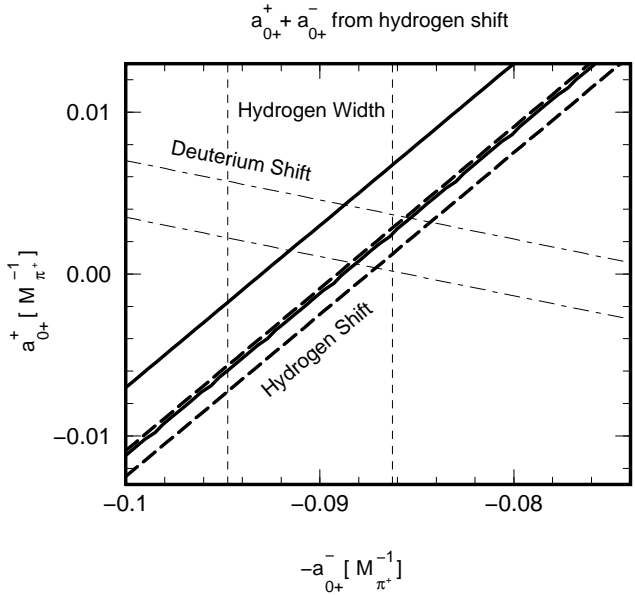}

\vspace*{1.1cm}

 \caption{\it
Determination of $\pi N$ scattering lengths from the pionic hydrogen and
pionic deuterium measurements. Solid line corresponds to the energy-level
shift calculations at $O(p^2)$ in ChPT, and the dashed lines - to the
potential model results
    \label{fig1}} 
\end{figure}

As it was mentioned in the Introduction, a new experiment at PSI is approved,
that is aimed at a precise determination of the $\pi^-p$ atom decay width.
At the first non-leading order in isospin breaking, the width is determined by
the charge-exchange amplitude $\pi^-p\rightarrow\pi^0n$ where the low-energy
constant $f_1$ does not appear. Consequently, the calculations for
this quantity can be carried out with a less theoretical uncertainty.

\section{Construction of the potentials from field theory}

The study of the properties of the hadronic atoms within the potential
approach has a decades-long history.
As a general rule, the predictions made within the potential model badly
deviate from the ones obtained within the field-theoretical approach based on
ChPT. For example, the potential model predicts the isospin-breaking 
correction $\delta_\Gamma$
to the $\pi^+\pi^-$ atom decay width with an opposite sign and with the same
order of magnitude~\cite{Rasche}. The predictions for the $\pi^-p$ atom
energy-level shift~\cite{PSI1} have been already considered in the previous
section. The reason for this discrepancy is now well understood. Namely, the
potential model does not take into account a full content of isospin-symmetry
breaking in QCD. In particular, the effect of the direct quark-photon
interactions (encoded in the ``electromagnetic'' low-energy constants of ChPT),
as well as the effect stemming from the dependence of scattering amplitudes on
the quark masses, are not included. As these effects contribute a bulk of the
total isospin-breaking correction term in the field-theoretical approach, 
it does not come to our surprise that the predictions of both approaches
substantially differ.

Since the isospin-breaking effects discussed above lead, in the language of
the potential scattering theory, to the modification of the short-range part
of the hadronic potential, it is evident that in order to bring the potential
model in conformity with the field-theoretical treatment, one has to assume
that short-range hadronic potentials contain an isospin-breaking piece.
It is natural to seek a derivation
of the potentials that are used in the potential model, on the basis of ChPT.
In a slightly more restricted context, one may ask, how the isospin-breaking
part of the short-range ``strong'' potential is obtained from ChPT, when
the isospin-symmetric part is already known to fit well ChPT predictions
(we recall that the isospin-breaking part is assumed to vanish identically in
existing potential models~\cite{PSI1,Rasche}).

It is widely presumed that the potential constructed from the field theory
will be necessarily singular in the position space and will require some
kind of regularization~\cite{potentials}. We argue that this is not
necessarily the case: almost any well-behaved
short-range potential, including those that were used in
Refs.~\cite{PSI1,Rasche}, can be generalized to include properly the full
content of isospin-breaking effects in ChPT.

The key observation that leads to the above conclusion, can be summarized in
the so-called universality conjecture.
This conjecture - completely in spirit of the
low-energy effective Lagrangian approach to bound systems - states that
the bound-state energies in the field theory, and in the potential model are
the same at the first order in isospin breaking, provided the threshold
amplitudes calculated in these two theories, coincide. We shall
ensure the universality for the case of a simple one-channel model where the
interaction Hamiltonian is given by a sum of Coulomb
and short-range interactions ${\bf H}_{\rm I}=-\alpha r^{-1}+{\bf U}$.
The short-range potential ${\bf U}$, in general, contains isospin-breaking
effects. 

With a given interaction potential, one may evaluate the energy-level shift
of the ground state of the bound system. The equation for the position
of the bound-state pole in the (complex) energy plane is given
by~\cite{Bern1}
\eq\label{pole-position}
&&z-E_0-\langle\Psi_0|\tau(z)|\Psi_0\rangle=0\, ,
\\[2mm]
&&\tau(z)={\bf U}+
{\bf U}(z-{\bf H}_{\rm 0})^{-1}(1-|\Psi_0\rangle\langle\Psi_0|)\tau(z)\, ,
\nonumber
\en
where $E_0$ denotes the unperturbed Coulomb energy of the ground state, and
$\Psi_0$ stands for the unperturbed wave function. The iterative solution of
Eq.~(\ref{pole-position}) yields usual Rayleigh-Schr\"{o}dinger perturbation
series. Further, it turns out useful to introduce the scattering matrix on the
short-range potential only
\eq\label{t}
{\bf t}(z)={\bf U}+{\bf U}\,\frac{1}{z-{\bf H}_0}\,{\bf t}(z)\, ,\quad\quad
{\bf t}_0\equiv\langle{\bf q}|{\bf t}(z)|{\bf p}\rangle\biggl|_
{{\bf p}={\bf q}=0,~z=0}\, .
\en
Here ${\bf t}_0$ denotes the value of the scattering amplitude at threshold
(for convenience, we have shifted the threshold to $z=0$).

The perturbative solution of the equation for the bound-state energy at
$O(\alpha^4)$ gives
\eq\label{level-shift}
\epsilon_{1s}=\frac{\alpha^3\mu_c^3}{\pi}\,
\biggl\{ {\bf t}_0-\frac{\alpha\mu_c^2}{\pi}\,{\bf t}_0^2\,
\biggl(1+\ln\,\frac{b}{2\alpha\mu_c}\biggr)
+\frac{8\alpha\mu_c}{\pi}\,{\cal Q}[{\bf t}]
+\frac{\alpha\mu_c^2}{\pi^3}\,{\cal R}[{\bf t};b]\biggr\}\, ,
\en 
where ${\cal Q}$ and ${\cal R}$ are certain known functionals of the
scattering matrix ${\bf t}$. We do not display the explicit expressions here.
The dependence on the arbitrary cutoff parameter $b$ in the functional 
${\cal R}$ cancels with the logarithmic dependence in the second term, so that
$d\epsilon_{1s}/db=0$.

At the next step, we consider the full scattering
amplitude defined through the Lippmann-Schwinger equation
${\bf T}(z)={\bf H}_{\rm I}+{\bf H}_{\rm I}(z-{\bf H}_0)^{-1}{\bf T}(z)$.
The threshold behavior of ${\bf T}(z)$ on energy shell is given by
(cf with Eq.~(\ref{threshold-piN}))
\eq\label{threshold}
&&{\rm Re}\,[{\rm e}^{-2i\theta_c}\langle{\bf p}|({\bf T}(z)-{\bf H}_I)
|{\bf q}\rangle]\biggl|_{|{\bf p}|=|{\bf q}|,~z={\bf p}^2/\mu_c}=
\frac{\tilde B_1}{|{\bf p}|}+\tilde B_2\ln\frac{|{\bf p}|}{\mu_c}
\nonumber\\[2mm]
&&\hspace*{4.cm}-\frac{2\pi}{\mu_c}\,{\cal A}+O(|{\bf p}|)+o(\alpha)
\\[2mm]
&&-\frac{2\pi}{\mu_c}\,{\cal A}=
{\bf t}_0-\frac{\alpha\mu_c^2}{\pi}\,{\bf t}_0^2\,\ln\,\frac{b}{2\mu_c}
+\frac{8\alpha\mu_c}{\pi}\,{\cal Q}[{\bf t}]
+\frac{\alpha\mu_c^2}{\pi^3}\,{\cal R}[{\bf t};b]+o(\alpha)\, ,
\en
with the same functionals ${\cal Q}$ and ${\cal R}$.

If one now expresses the energy level shift in terms of the threshold
amplitude ${\cal A}$, one arrives at exactly the same
expression~(\ref{general-piN}) as in the field-theoretical framework -
in accordance with the universality conjecture
\eq\label{energy}
\epsilon_{1s}=-2\alpha^3\mu_c^2{\cal A}\,(1-2\alpha\mu_c(\ln\alpha-1)
{\cal A})+o(\alpha^4)\, . 
\en

Based on the universality conjecture, we can provide a constructive algorithm
for the derivation of the isospin-breaking part of the short-range
potential ${\bf U}$ from ChPT. The amplitude at threshold in the latter
is generally given by ${\cal A}={\cal A}_0+{\cal A}_1+\cdots$, where
${\cal A}_{0},~{\cal A}_1$ 
denote the isospin-conserving (breaking) parts of the 
amplitude, and ellipses stand for higher-order terms in isospin breaking.
In order to ensure the inclusion of the full content of isospin-symmetry
breaking in ChPT into the potential model, it thus suffices to match the
amplitude ${\cal A}$ in both theories. The problem evidently has too much
degrees of freedom. The short-range potential is also given by the sum of
isospin-conserving and isospin-breaking pieces 
${\bf U}={\bf U}_0+{\bf U}_1+\cdots$, and for our purposes the following 
{\it ansatz} is sufficient: ${\bf U}=(1+\lambda){\bf U}_0$, where the sole
coupling $\lambda$ will be determined from matching the isospin-breaking
pieces of the threshold amplitude.
 
If one defines the scattering amplitude in the limit of no isospin breaking
through $\bar{\bf t}(z)={\bf U}_0+{\bf U}_0(z-{\bf H}_0)^{-1}\bar{\bf t}(z)$,
The matching condition may be written recursively, in terms of $\bar{\bf t}$
\eq\label{match}
&&-\frac{2\pi}{\mu_c}\,{\cal A}_0=\bar{\bf t}_0\, ,
\\[2mm]
&&-\frac{2\pi}{\mu_c}\,{\cal A}_1=\lambda(\bar{\bf t}_0+{\cal S}[\bar{\bf t}])
-\frac{\alpha\mu_c^2}{\pi}\,\bar{\bf t}_0^{~2}\,\ln\,\frac{b}{2\mu_c}
+\frac{8\alpha\mu_c}{\pi}\,{\cal Q}[\bar{\bf t}]
+\frac{\alpha\mu_c^2}{\pi^3}\,{\cal R}[\bar{\bf t};b]\, .
\nonumber
\en
where ${\cal S}$ again stands for a certain known functional.
The matching condition~(\ref{match}) solves our problem completely -
the bound-state energies calculated with the use of the ``corrected''
potential coincide, by definition, with those calculated on the basis of
ChPT.
\section{Conclusions}

The approach based on non-relativistic effective Lagrangian technique and ChPT,
provides one a powerful tool to systematically calculate the characteristics
of loosely bound states of hadrons.
With the use of this approach, the $\pi^+\pi^-$ atom decay problem is now 
completely understood, both conceptually and numerically.
Certain theoretical effort will be still needed to extract the precise values 
of $\pi N$ scattering lengths from past and future measurements at PSI.
The treatment of other systems, like kaonic atoms that will be measured by the
DEAR experiment at DA$\Phi$NE, is foreseen within the same framework.

Further, within the present approach, one may establish a constructive 
algorithm for the derivation of the short-range hadronic potentials from ChPT.
The algorithm is based on the universality conjecture that has been discussed
above, for a simple one-channel case.
We hope that - after the suitable generalization - the approach based
on the universality might be also useful for the analysis of $\pi N$ 
scattering data near threshold, in what concerns the study of the 
isospin-breaking effects in the $\pi N$ amplitude.

\vspace*{.2cm}

{\it Acknowledgments}. 
This work was supported in part by the Swiss National Science
Foundation, and by TMR, BBW-Contract No. 97.0131  and  EC-Contract
~No. ERBFMRX-CT980169 (EURODA$\Phi$NE).

\end{document}